\PassOptionsToPackage{warn}{textcomp}
\documentclass[sigconf]{acmart} 

\def\BibTeX{{\rm B\kern-.05em{\sc i\kern-.025em b}\kern-.08emT\kern-.1667em\lower.7ex\hbox{E}\kern-.125emX}}

\usepackage{lipsum}
\usepackage{textcomp}
\usepackage{times}
\usepackage{float}
\usepackage{booktabs} 
\usepackage{color,soul}
\usepackage{xcolor}
\usepackage{multirow}
\usepackage{amsmath,amssymb}
\usepackage{seqsplit}

\newif\ifnotes
\notestrue 

\ifnotes
\newcommand{\inred}[1]{{\color{red}\sf\textbf{\textsc{#1}}}}
\newcommand{\note}[1]{\frameit{Note}{#1}}
\newcommand{\todo}[1]{\frameit{To-do}{#1}}
\newcommand{\inote}[1]{\inred{$<<<${#1}$>>>$}}
\else
\newcommand{\inred}[1]{#1}
\newcommand{\note}[1]{}
\newcommand{\todo}[1]{}
\newcommand{\inote}[1]{}
\fi

\newcommand{\remove}[1]{}

\def \review #1{\color{black}#1\color{black}}

\newcommand{\etal}{\textit{et al}. }

\DeclareMathOperator*{\argmax}{arg\,max}
\DeclareMathOperator*{\argmin}{arg\,min}

\setcopyright{rightsretained}

\copyrightyear{2019}
\acmYear{2019}
\setcopyright{acmlicensed}
\acmConference[WiSec '19]{12th ACM Conference on Security and Privacy in Wireless and Mobile Networks}{May 15--17, 2019}{Miami, FL, USA}
\acmBooktitle{12th ACM Conference on Security and Privacy in Wireless and Mobile Networks (WiSec '19), May 15--17, 2019, Miami, FL, USA}
\acmPrice{15.00}
\acmDOI{10.1145/3317549.3323411}
\acmISBN{978-1-4503-6726-4/19/05}

\begin{document}

\title{Deployment Optimization of IoT Devices through Attack Graph Analysis}



\author{Noga Agmon, Asaf Shabtai, Rami Puzis}

\affiliation{
 \institution{Department of Software and Information Systems Engineering,\\ 
 Ben-Gurion University of the Negev\\
 nogaag@post.bgu.ac.il, \{shabtaia, puzis\}@bgu.ac.il
 }
}

 

\begin{abstract}
\review{
The Internet of things (IoT) has become an integral part of our life at both work and home. 
However, these IoT devices are prone to vulnerability exploits due to their low cost, low resources, the diversity of vendors, and proprietary firmware. 
Moreover, short range communication protocols (e.g., Bluetooth or ZigBee) open additional opportunities for the lateral movement of an attacker within an organization. 
Thus, the type and location of IoT devices may significantly change the level of network security
of the organizational network.
In this paper, we quantify the level of network security based on an augmented attack graph analysis that accounts for the physical location of IoT devices and their communication capabilities.
We use the depth-first branch and bound (DFBnB) heuristic search algorithm to solve two optimization problems: Full Deployment with Minimal Risk (FDMR) and Maximal Utility without Risk Deterioration (MURD).
An admissible heuristic is proposed to accelerate the search. 
The proposed method is evaluated using a real network with simulated deployment of IoT devices. The results demonstrate (1) the contribution of the augmented attack graphs to quantifying the impact of IoT devices deployed within the organization on security, and (2) the effectiveness of the optimized IoT deployment. }

\end{abstract}

%
%

\ccsdesc[500]{Security and privacy~Distributed systems security}
\ccsdesc[500]{Security and privacy~Mobile and wireless security}

\keywords{Attack graphs, Internet of Things, IoT deployment, Optimization, Short-Range Communication}

\maketitle


\section{Introduction}
It is estimated that by 2020 more than 20 billion IoT devices will be deployed in the world ~\citep{Meulen2017Gartner2016}. 
Most IoT products are not equipped to deal with security and privacy risks, which can turn them into the weakest link of organizational networks. 
The risk of IoT devices to the security of an organization is underestimated in many cases when an organization's IT department relies solely on network separation to isolate IoT devices from other IT assets. 
Such an approach disregards some of the unique properties of IoT devices, such as light or sound emissions, various sensors, and diverse communication protocols such as NFC, Bluetooth, ZigBee and LoRA, in addition to standard Wi-Fi. 
The advanced capabilities of IoT devices can be exploited by an attacker for lateral movement within an organization, shoulder surfing, and more, making them a valuable asset for an attacker.

With respect to hardening IoT security, most prior research focuses on the security of individual IoT devices~\citep{roman2011securing, liu2012research, zhang2014iot}, the security of an IoT protocol~\citep{Vaccari2017RemotelyNetworks, Zillner2015ZigBeeUgly, morgner2016all, wright2009killerbee, ronen2017iot}, or the the security of a network that consists solely of IoT devices~\citep{huang2014novel, skarmeta2014decentralized, zanella2014internet, Ge2016} (see Section~\ref{sec:rw_iot_dep} for more details). \review{To the best our knowledge, there is no previous related research aimed at identifying the optimal (security risk-wise) deployment of devices within the physical space. }
The location of an IoT device within an organization can have unintended effects on the network topology such as bridging between networks through short-range communication protocols (see Sections~\ref{sec:back-iot} and \ref{sec:back-src}). 
We use the following example to demonstrate the problem.

\begin{figure}
\centering
\includegraphics[width=0.45\textwidth]
{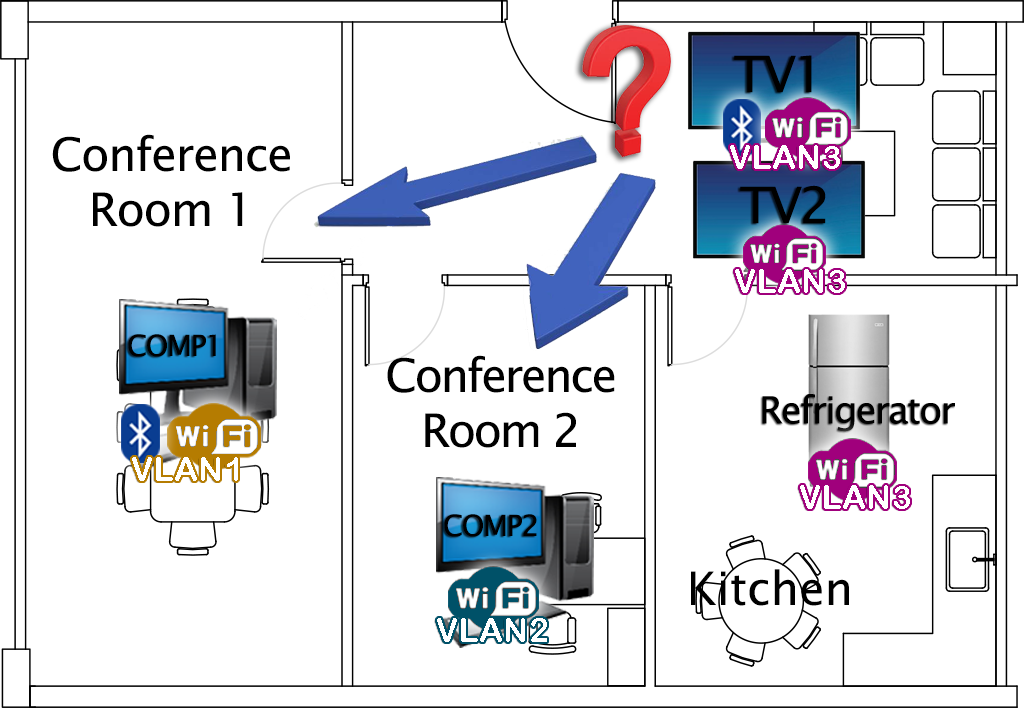}
\caption{Illustration of the office in example~\ref{run_example}. Different colors of Wi-Fi represent different VLANs.} 
\label{fig:example_setup}
\end{figure}

\begin{example}
\label{run_example}
Assume, for example, an office with two conference rooms and a kitchen (Figure~\ref{fig:example_setup}).
Each conference room has a computer ($COMP1$ and $COMP2$) connected through Wi-Fi to two different VLANs ($VLAN1$ and $VLAN2$ respectively). $COMP1$ also has Bluetooth. A smart refrigerator in the kitchen is connected to $VLAN3$ and has Internet connectivity. All other IoT devices in the office are connected to $VLAN3$ as well. The office purchased two televisions ($TV1$ and $TV2$) to replace the old \review{projectors } in the conference rooms. Both televisions are connected to $VLAN3$ via Wi-Fi; 
TV1 is also equipped with Bluetooth.

\textbf{Should we install $TV1$ in Conference Room 1 and $TV2$ in Conference Room 2 or vice versa?} 
To answer this question assume, for example, that unbeknownst to the organization, a sophisticated malware has managed to infect one of the computers in the organizational network.
Further, assume that the malware is equipped with the necessary exploits to hop between devices in the office. If $TV1$ is placed in Conference Room 1, the attacker could take advantage of the fact that both $TV1$ and $COMP1$ have Bluetooth and create an attack path to the refrigerator. However, if $TV1$ is placed in Conference
Room 2 this attack path will no longer be available to the attacker.
\end{example}

The risk of potential multi-step attacks such the one described in Example~\ref{run_example}  can be estimated using attack graphs~\citep{phillips1998graph,ou2006scalable}.
An attack graph is a model of a computer network that encompasses computer connectivity, vulnerabilities, assets, and exploits. It is used to represent a collection of complex multi-step attack paths (hereafter referred to as \emph{attack plans}) and can be used to assess and quantify security risk (see Section~\ref{sec:back-ag} for more details). 

In this paper, the proposed method augments attack graph analysis to account for the physical location of IoT devices and their communication capabilities. (see Section~\ref{sec:iot_ag}).
Relying on the new attack graphs, we quantify the risk of adding an IoT device to a given network and show that the number of short attack paths may increase by 19\% due to the deployment of only six IoT devices in a small to medium sized enterprise; short attack plans often pose the greatest threat, because they represent an attack that needs fewer resources to be executed.

We also optimize the deployment of IoT devices in order to reduce the negative security implications of such deployment (see Section~\ref{sec:deployment_opt}). 
Two optimization problems are presented: 
the Full Deployment with Minimal Risk (FDMR) problem where all required IoT devices should be deployed with minimal security implications and the Maximal Utility without Risk Deterioration (MURD) problem where the maximal number of IoT devices should be deployed without increasing the security risk of the network. 
We use depth-first branch and bound (DFBnB) heuristic search algorithm to solve both optimization problems and suggest an admissible heuristic function to accelerate the search.
Our experiments show that optimal deployment of IoT devices can reduce the number of possible attack plans by 18\% (see Section~\ref{sec:evaluation}).

\section{Background} 
\label{sec:background} 
\subsection{Attack Graphs}
\label{sec:back-ag}

An attack graph is a model of a computer network that encompasses computer connectivity, vulnerabilities, assets, and exploits~\citep{phillips1998graph,ou2006scalable}. 
Attack graphs are used to represent collections of complex multi-step attack scenarios traversing an organization from an initial entry point to the most critical assets. 
By analyzing the attack graph, a security analyst can assess the risks of potential intrusions and devise effective protective strategies. 
The attack graph analysis methodology contains three main stages: (1) network and vulnerability scanning, (2) attack graph modeling, and (3) attack graph analysis. 

In the first stage, the Nessus vulnerability scanner~\citep{beale2004nessus} is used in order to map the vulnerabilities of all of the hosts in the organization. 
Connectivity between the hosts can be identified manually by system administrators based on the organizational network topology and firewall configurations. 
Nessus, Nmap, or other network scanners can aid in the connectivity assessment process. 

Network connectivity and vulnerability reports are processed by MulVAL~\citep{ou2005mulval} to generate an attack graph representation in planning domain definition language (PDDL). 
\review{An attack graph consists of privilege nodes, exploit/action nodes, and fact nodes. In an attack graph, a privilege node represents the information gained or the access privileges that the attacker obtains (represented by triangles in the graph). An exploit/action node represents the action the attacker needs to exploit a vulnerability (represented by ovals). The edges of exploit nodes are for preconditions and postconditions of the exploit. A fact node represents a network condition that needs to exist in order for the attacker to exploit the vulnerability (represented by rectangles).
To gain a privilege, an attacker needs to execute one of the actions leading to it (logical OR). To use an exploit, the attacker needs all of the privileges and the facts that lead to the exploit (Logical AND). An exploit node needs all of these preconditions leading to it to be executed, and once executed, the attacker gains all of the postconditions the exploit node leads to~\citep{ou2006scalable,Sawilla2008IdentifyingGraphs,Ammann2002ScalableAnalysis,noel2014metrics}.}

\begin{figure}
\centering
\includegraphics[width=0.45\textwidth]
{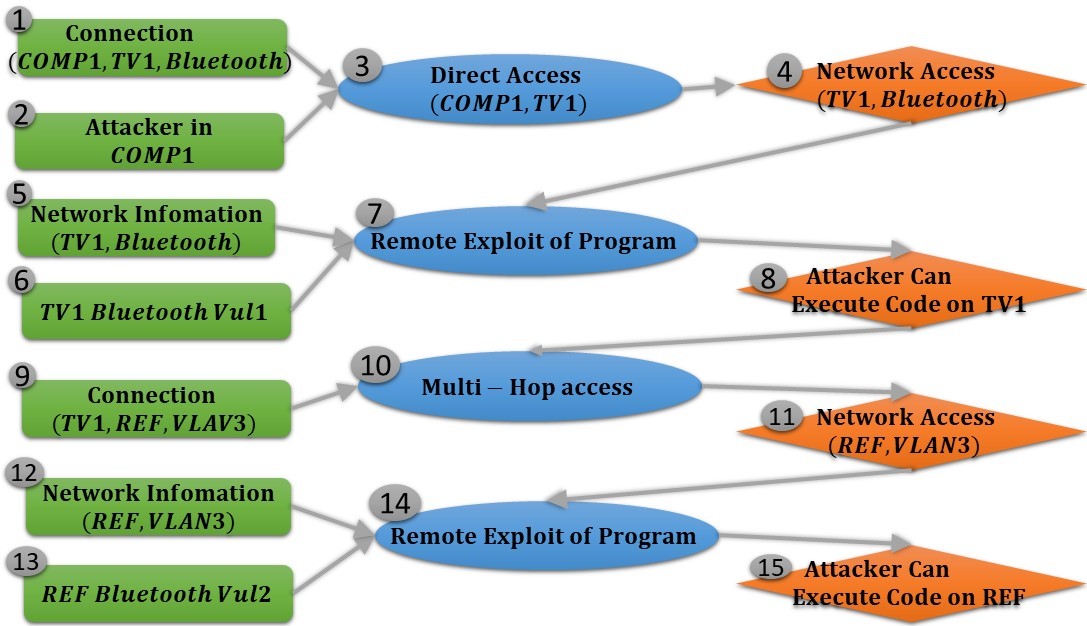}
\caption{Attack Graph of Example~\ref{run_example_ag}. \review{ Exploit/action nodes are represented by blue ovals; fact nodes are represented by green rectangles; privilege nodes are represented by orange diamonds.}} 
\label{fig:attack_graph}
\end{figure} 

\begin{example}
\label{run_example_ag}
Figure~\ref{fig:attack_graph} presents an abstract attack graph of the situation described in example~\ref{run_example}.
At the top of the figure, two fact nodes (nodes 1 and 2) that represent two facts of the system can be seen (green rectangles).
Access between $COMP1$ and $TV1$ can only created if these two conditions exist, as can be seen from the blue oval, which represents an exploit node (node 3). 
This access allows the attacker to use the Bluetooth connectivity of $TV1$, as represented by the orange diamond (node 4), meaning that the attacker can obtain control of $TV1$ via $COMP1$.
\end{example}

Following the construction of an attack graph, the graph's PDDL representation can be used as a domain model for variety of planners. 
A typical task is finding the optimal attack plan or estimating the likelihood of a successful attack given the attack graph of an organization~\cite{Singhal2011SecurityGraphs, Wang2008AnMetric, noel2014metrics}.
Consequently, attack graphs can be used for hardening network security through a variety of attack graph optimizations~\cite{islam2008heuristic, abadi2006ant, polad2017attack, noel2003efficient}.

\subsection{Security in the Internet of Things}
\label{sec:back-iot}
Traditional security solutions such as firewalls, IDSs, anti-viruses, and software patches are not suitable for IoT devices.
The three major reasons for this are~\citep{Yu2015HandlingDevices}: 
(1)~Types of policies: a single app may use several IoT devices, communicating explicitly (e.g., via Wi-Fi or Bluetooth) or implicitly (e.g., an IoT light bulb can be triggered by an IoT light sensor). 
The outcome is a complex and dynamic network which can be hard to secure using a single security policy \review{(e.g., with firewalls)}. 
(2)~Signatures and anomalous behavior recognition: some security methods store anomalies and signatures on the device to recognize and detect threats. 
Due to the diversity of IoT devices and manufacturers, these methods will be inadequate, \review{mainly because of the constant need to update and maintain the device to support these tools. }
(3)~Enforcement mechanism: IoT devices have low computation abilities, low power consumption, and do not run full-fledged operating systems. Most common security methods need all of the above to operate and therefore are impractical to implement on IoT devices.
\review{(4)~Unsupported devices: the longevity of IoT devices can lead to deployed devices that vendors no longer support. In that way, vulnerable devices (with default passwords or unpatched bugs) can remain in the organization. }

Moreover, the competitive IoT device market compels vendors to try and get their products out as fast as they can, prioritizing functionality and the user experience, and ignoring the security aspect. 
In general, most products hardly deal with security and privacy risks, making them the weakest link in terms of security and the target of attackers interested in breaking into networks and harming systems or leaking information~\citep{Yu2015HandlingDevices}. 
Thus, despite the fact that security was recognized as a central issue of the IoT market as early as 2011 by Bandyopadhyay \etal~\citep{Bandyopadhyay2011InternetStandardization}, it still continues to remain a challenge today. 

\subsection{Short Range Communication Protocols}
\label{sec:back-src}
When connecting a device to a network it is possible to use two categories of networking technologies. 
The first and simplest category is to connect using standard existing network technologies such as Wi-Fi and Ethernet. The second category is to connect using different wireless technologies that are more suitable for some devices, e.g., technologies that are more appropriate for devices that require low energy consumption protocols.
These protocols are short-range communication protocols, due to their requirement for short proximity in order to perform a connection.

Currently, in the second category there are several communication methods that can be used, including: ZigBee, Z-Wave, Powerline, Bluetooth 4.0, and other radio frequency protocols, but no standard protocol exists. Both Z-Wave and ZigBee are considered secure, but implementation flaws and manufacturer mistakes make them vulnerable~\citep{InsecuritySymantec}. 

In our research, we focus on ZigBee \review{and Bluetooth, since they are ones } of the most common wireless technologies used to connect IoT devices. 
\review{First, we start with the ZigBee protocol, } 
which guarantees low power consumption and a two-way, reliable, wireless communications standard for short-range applications. 
It is open-source and has advantages such as easy deployment and global usage.

The ZigBee protocol was created with security considerations in mind, but consumer demand for cheap devices with long life expectancy often caused vendors to sacrifice security, which led to poor implementation of the protocol~\citep{Zillner2015ZigBeeUgly}; this, in turn, led to major security issues such as data compromising or information sniffing.~\citep{Vaccari2017RemotelyNetworks}.
For example, Vaccari \etal~\citep{Vaccari2017RemotelyNetworks} focused on the security aspects of the ZigBee protocol. 
The study identified important security issues and presented an attack on the protocol which enabled the attacker to compromise the data transferring in the network.  
Morgner \etal~\citep{morgner2016all} described a novel attack that shows that the ZigBee Light Link standard is insecure by design. 
Wright \etal~\citep{wright2009killerbee} published KillerBee, a penetration testing tool which allows ZigBee traffic to be sniffed and analyzed.
Ronen \etal~\citep{ronen2017iot} found a major bug in the ZigBee protocol in Philip Hue smart lamps. 
They were able to perform an over-the-air firmware update, thereby infecting the lamp with a worm that can spread to any of the lamp's neighbors.

\review{
Bluetooth was developed by a group called the Bluetooth Special Interest Group (SIG) in May 1998. Today, a lot of smartphones, sports devices, sensors, and medical devices have Bluetooth. 
The protocol become widely used because of its low cost and low power consumption.}

\review{
In~\citep{ryan2013bluetooth}, techniques were presented for eavesdropping on devices using Bluetooth. 
An extended review of Bluetooth threats and possible attacks was performed by Minar~\etal, Sandya~\etal and Dunnin ~\citep{minar2012bluetooth, sandhya2012analysis, dunning2010taming}, and recently, Cope \etal~\citep{cope2017investigation} investigated the currently available tools to exploit vulnerabilities in Bluetooth. 
In conclusion, many Bluetooth versions that are in use today, have a wide variety of security vulnerabilities.
}

In addition to the security issues, the number of communication protocols in an IoT device can also influence the security of the device. 
If such a device is compromised by an attacker that has hacked into one of its communication protocols, the hacker can take advantage of the compromoised device and use the other protocols as entry points to the network~\citep{Ge2017SecurityDevices}.

Of all the above, short-range communication protocols are another aspect of IoT devices that make them insecure compare to regular~hosts.

\subsection{Heuristic Search }
\label{sec:back-search}
Heuristic search is a family of techniques used to solve difficult problems in artificial intelligence (AI). 
In this case, each problem is represented by states, where each state represents the current condition of the problem. 
Each problem also has a starting state and one or more goal states. 
A search space is the environment in which a search takes place, where the purpose of the search is to find a path from the start state to one of the goal states in the search space. 
Each solution represents by one goal state. 
The quality of the solution is measured by the cost of the goal state.
Search algorithms make a distinction between minimum and maximum problems. 
In a minimum problem, we want to find the solution with the lowest cost, and in a maximum problem the highest cost solution is desired. 
Most problems are minimum problems, e.g., we want the cheapest or the fastest solution. If not stated differently in this paper, we are referring to a minimum problem.
In our research, we use the depth-first branch and bound algorithm~\citep{Korf2010ArtificialAlgorithms,Zhou2006Breadth} which uses a heuristic function to solve the problems more efficiently.

\subsubsection{Heuristic}
A heuristic is an estimation of the cost of the path from node $n$ to a goal node. The heuristic function is used to steer the search algorithm in the direction of the goal.
In an informed way, heuristics help the algorithm guess which child out of all of the node's children will lead to the goal.

\textbf{Admissible Heuristic. } 
If, for any $n$, a heuristic function never overestimates the cost of the best path from node $n$ to a goal node, then the function is referred to as an \emph{admissible heuristic function}. 
Note that in a maximum problem (where we want the solution with the maximum cost) it is the opposite, i.e., a heuristic function that never underestimates the cost of the best path. 


In most search algorithms, one of the most important conditions for a heuristic function is that it should be admissible.


\subsubsection{DFBnB Algorithm}
depth-first branch and bound (DFBnB) is a depth-first search algorithm~\citep{Zhou2006Breadth, Korf2010ArtificialAlgorithms}.
The algorithm is used to navigate through the search space and find the optimal solution.
During the search process, DFBnB maintains the best solution found so far. 
In order to perform pruning more frequently and thus accelerate the search process, DFBnB uses a heuristic function.
The algorithm returns an optimal solution with linear memory space, assuming the heuristic function is admissible.

DFBnB prunes subtrees of the search space whenever the algorithm can prove that no solution can be found that is better than the incumbent solution. 
This solution depends on the kind of problem (i.e., minimum or maximum),
which is determined by the cost of the goal state.
\remove{
The cost of a node is defined by $f(x) = g(x)+h(x)$, where f(x) is the sum of (1) the current cost of the node $g(x)$, and (2) the heuristic function for the goal $h(x)$, which is the estimation of the cost from node $x$ to the goal. 
Thus, $f(x)$ estimates the lowest total cost of any solution path going through node $x$.
If a branch with a higher cost is found, it can be pruned so that there is no need to keep expanding towards it.
The solution of the problem is determined by the cost of the goal state.
}



\review{\section{Related Work}} 
\label{sec:related_work}
\subsection{IoT Device Deployment}
\label{sec:rw_iot_dep}
There are several works regarding the deployment of IoT devices, but most of them do not consider the security aspect. 
For example, Huang \etal~\citep{huang2014novel} proposed a deployment scheme used to a achieve green networked IoT, while Skarmeta \etal~\citep{skarmeta2014decentralized} focused on privacy issues and Zanella \etal~\citep{zanella2014internet} focused on the IoT in smart cities.

Some of the research that refers to security analyzes single IoT devices but does not look at IoT devices as a deployment problem.
Liu \etal~\citep{liu2012research} tried to solve the problem of assessing the risk of a single IoT device, by proposing a dynamical risk assessment method inspired by an artificial immune system. 
Zhang \etal~\citep{zhang2014iot} and Roman \etal~\citep{ roman2011securing} reviewed security issues in the IoT in terms of the security of each device. 

There are a few works that refer to deployment and network security, but they do not take the combination of hosts (such as computers and servers) with IoT devices into consideration.
Mohsin \etal~\citep{mohsin2017iotriskanalyzer} argued that the likelihood of exploiting IoT vulnerabilities depends on the system configuration. 
The authors explained that various configurations derive from different devices, technologies, and connectivity, all of which serves the same goal but have different risk levels.
Santoso \etal~\citep{santoso2015securing} presented an approach to secure smart home systems in which IoT devices are deployed, and Abie \etal~\citep{abie2012risk} introduced a risk-based adaptive security framework for the IoT in health-care systems.
The research mentioned above reflects the many challenges of IoT security. 
In this respect, our work is unique in two ways. 
First, it combines the security concerns of the IoT with workstations and servers, while taking into account the possible use of one to hack the other.  
Second, our network model is a generic network that \review{can be suitable for a variety of scenarios } and is not specific for a particular domain.

\subsection{Attack Graph Optimization}
\label{sec:rw_ag-opt}

\textbf{Attack Graph Representation. } 
Attack graphs have been used to estimate the security risk score of organizational networks~\citep{Singhal2011SecurityGraphs, Wang2008AnMetric, noel2014metrics},
however the specific characteristics of IoT devices were not considered in these articles.
In all of this research, the structure of the regular IT network is analyzed, taking into account the vulnerabilities of workstations and servers. 
IoT devices introduce additional challenges to security risk modeling through attack graphs, such as the diverse physical locations, variety of short-range communication protocols, cyber-physical capabilities of the devices, mobility, etc. 

In this paper, we augmented the attack graph model of an organization to consider locations and short-range communication of IoT devices, and we used the augmented attack graph model to optimize the deployment of IoT devices throughout the organization. 

\textbf{Risk Score. } 
\review{Wang \etal~\citep{Wang2008AnMetric} suggested an overall network security score by combining individuals' vulnerabilities regarding their relationship in attack graphs.
Singhal \etal~\citep{Singhal2011SecurityGraphs} defined the risk score as the likelihood of an attack which was derived from the likelihood of individual exploits. 
Noel \etal~\citep{noel2014metrics} described four families of metrics for measuring security risk in attacks graph. 
Every family was represented by one entry in a four-dimensional vector.   
The Euclidean norm of this vector was used as the overall risk score. }
\review{Gonda \etal~\citep{gonda2017ranking} computed the number of shortest plans in a planning graph derived from an attack graph as a way to measure the security of the network, and Swiler \etal~\citep{swiler2001computer} computed the set of near-optimal shortest paths to identify the most exploitable components in the network. Polad \etal~\citep{polad2017attack} used an attack graph to estimate the security of the network as the cost of the attack path that led to the goal. } 

\review{All the above risk scores can be used to optimize the IoT deployment once the attack graph definition has been augmented to take into account the IoT device specifications. 
In this paper, we adopt Gonda's approach to measure network security and combined it with Polad's method, to include the length of the shortest plans, as well as their quantity (see Section~\ref{sec:risk_score}). }

\review{
\textbf{Optimization Problems.} 
Security risks can be reduced by patching vulnerabilities. 
However, it is not always possible to patch all vulnerabilities at once due to operational costs (patching often requires significant downtime). 
A variety of low cost network hardening approaches can be used to prioritize the vulnerabilities (e.g., \citep{noel2003efficient,jun2011minimum})}. 
\review{Islam \etal~\citep{islam2008heuristic} argued that most of these methods are not scalable. 
They proposed heuristic algorithms to accelerate the patch optimization. Abadi \etal~\citep{abadi2006ant} used the ant colony optimization algorithm to detect a minimum critical set of exploits. }
\review{Polad \etal~\cite{polad2017attack} examined the effect of adding fake vulnerabilities in an attack graph and used combinatorial optimization in order to find optimal assignment of these vulnerabilities.
Almohri \etal~\citep{almohri2016security} used sequential linear programming in attack graphs to find the optimal placement of security products (e.g., a host-based firewall) across a network. 
The authors used a probabilistic model which uses Bernoulli and transformed the attack graph into a system of linear and nonlinear equations.
Noel \etal~\citep{noel2008optimal} used attack graph to optimize the placement of intrusion detection system (IDS) sensors to allow monitoring malicious activity on critical paths. }

\review{In this paper we present a different optimization problem of optimizing the set of IoT devices to be deployed throughout an organization with minimal implications to the network security. }

\subsection{IoT in Attack Graphs}
\label{sec:rw_iot_ag}
Very little work has been performed on attack graphs that consist of IoT devices. 
The first research performed in this area was conducted by Ge \etal~\citep{Ge2016} who used attack graphs in conjunction with IoT devices.
However, the network used consisted only of IoT devices, most of which were the same kind of device. 
The network topology was fixed, small, and relatively uncomplicated. 
The authors proposed a framework for IoT device security modeling with the aim of presenting all possible attack paths in the network, evaluating the security level, and assessing the effectiveness of different defense strategies.

In a later work, Ge \etal~\citep{Ge2017SecurityDevices} noted that some IoT devices use more than one communication protocol. 
The writers argued that if such a device is compromised by hacking into one of the communication protocols, the hacker can take advantage of it and use the other protocols as entry points to the network. 
The paper used HARMs (hierarchical attack representation models), which are models of attack graphs used, to improve scalability~\citep{hong2012harms}.
The authors presented a real scenario and showed how an attacker can take advantage of it. 
In the scenario, some devices have both Wi-Fi and ZigBee communication protocols. 
Also present are smart devices such as a tablet and TV that can connect to a Philips Hue lighting system (Hue Bridge) by Wi-Fi. 
This lighting system also has ZigBee which allows it to control smart light bulbs in the house.
By exploiting the tablet that runs the Hue application, an attacker can gain control of the Hue Bridge system and use it to control all of the smart lights. 
The authors noted that the lighting hub can consist of any other smart hub, and the scenario can also be used to hack into any smart device, not only light bulbs.

Yi{\u{g}}it \etal~\citep{yiugit2019cost} proposed COBANOT, a heuristic-based cost and budget aware network hardening solution for IoT systems which uses compact attack graphs~\citep{chen2010scalable}.
This work is the first to use attack graphs in IoT systems for network hardening. 
However, their experiment included a small-scale attack graph that only consists of IoT devices. In addition, none of the unique characteristics of IoT devices, such as different protocols, mobility, physical proximity, etc. were considered.

Our research focuses on networks that combine all kinds of hosts and IoT devices. 
Also, our network's size is larger than the networks used in the research mentioned above.



\section{IoT Attack Graphs}
\label{sec:iot_ag}
\subsection{IoT Deployment}
In a typical organization, all hosts (workstations and servers) are connected to the organizational network via a wired or wireless connection. 
Let $H = \{ h_1, h_2,\ldots, h_z\}$ be the set of hosts that are part of the organization network.

In addition to the regular hosts, the organizational network may contain IoT devices.
Let $D = \{d_1, d_2,\ldots, d_m\}$ indicates the set of unique IoT devices.
Each IoT device $d_i$ has a unique identifier (usually an IP address).

IoT devices differ by their purpose and capabilities. 
For example, a refrigerator is capable of maintaining a low temperature while a smart TV is capable of showing high definition movies. 
We group IoT devices by type, e.g., refrigerator,  TV, camera, smoke detector, etc.
$T = \{t_1, t_2,\ldots, t_n\}$ is the set of all the IoT device types. 
We denote a set of all devices that are of type $t$ as $D(t)$ and a single device type $d$ as $t(d)$.
We assume that every IoT device is part of just one group.

Some IoT devices can only be deployed in specific predefined designated locations. 
For example, the kitchen is typically the designated location for a refrigerator, while large TV screens or projectors are found in meeting rooms. 
Some IoT devices such as cameras or smoke detectors may be deployed in many different locations throughout an organization.
\begin{definition}[Locations]
\label{def:locations}
$L = \{l_1, l_2,\ldots, l_b\}$ indicates the set of unique location spots where IoT devices can be deployed. 
We denote the set of locations where an IoT device of a specific type $t\in T$ can be deployed as $L(t)\subseteq L$. 
\review{In every location spot only one type of IoT devices can be deployed, meaning, $L(t)$ is defined such that the intersection of each pair of $L(t_i)$ sets are empty, $\cap_{t \in L(t)} = \emptyset$. Because a location spot must be associated with some type of IoT devices, the union of $L(t)$ is equal to $L$, $\cup_{t \in L(t)} = L$}
\end{definition}

Organizations may have constraints about the deployment of IoT devices. 
We defined two main constraints for a device type $t$. 
The first one is the number of locations (out of the total locations available) that need to contain a deployed device of that type.
For instance, there are four possible locations for cameras in the hallway, but the organization only needs to deploy two of them.
The second constraint is the number of devices there are of each type. 
For instance, for one location in which a refrigerator can be deployed, there are three possible refrigerators that the organization can purchase.

\begin{definition}[Location Constraint]
\label{def:locations_const}

Let $C$ be the set of all constraints. $C(t)$ is a three-tuple that represents a constraint for a type $t$, $C(t) = (L(t), n(t), D(t))$. 

$L(t)$ is the set of locations that an IoT device of a specific type $t$ $\in$ $T$ can be deployed (as defined in Definition~\ref{def:locations}).

$n(t)$ is the number of locations that needed to be deployed out of all locations in $L(t)$.

$D(t)$ is a set of all IoT devices that are of type $t$.
\end{definition}

An example of a constraint can be derived from example~\ref{run_example}. 
Suppose the organization has three possible locations in which a TV can be deployed ($L(TV)=\{l_{TV_1},l_{TV_2},l_{TV_3}\}$) but only needs to deploy a TV in two of these locations ($n(TV)=2$). 
In addition, there are four different televisions that can be deployed ($D(TV)=\{d_{tv1},d_{tv2},d_{tv3},d_{tv4}\}$). 
Formally, constraint $C(TV)$ would be defined as follow: 

$C(TV) = (\{l_{TV_1},l_{TV_2},l_{TV_3}\},2, \{d_{tv1},d_{tv2},d_{tv3},d_{tv4}\})$.

Assume that at most one IoT device can be deployed in each location $l\in L$.
The deployment of IoT devices is defined as a function $depl:D\rightarrow L\cup \{\bot\}$ which maps every device to a particular location. 
The special non-location symbol $\bot$ signifies that a device is not deployed.  
We say that a deployment is valid if it does not violate the constraints specified in Definition~\ref{def:locations_const}.
\begin{definition}[Valid Deployment]
Let $depl:D\rightarrow L\cup \{\bot\}$ be a deployment of IoT devices. $depl$ is valid if $\forall_{d\in D}, depl(d)\in L(t(d))\cup\{\bot\}$.
\end{definition}

\review{We denote $depl_{full}$ as a deployment that satisfies all constraints~$C$ and $depl_{empty}$ as an empty deployment with no IoT devices deployed. }
Note that the condition $n(t) \le |D(t)|$ should be satisfied for full deployment to exist.

Many IoT devices deployed within an organization's premises will likely be able to communicate with nearby hosts via short-range communication (SRC) protocols such as ZigBee, Bluetooth, ad hoc Wi-Fi, etc. 
Some hosts within the organization may also support SRC protocols, which could allow the adversary to hop between networks.
\begin{definition} [Short-Range Communication]
We define a set of short-range communication protocols $SRC = \{p_1, p_2,\ldots\}$. 
Let $src:D\cup H\rightarrow 2^{SRC}$ be a function that maps an IoT device or a host to the subset of SRC protocols that it supports. 
\end{definition}
In the remainder of this paper we will use the term \emph{device} to refer to both IoT devices and hosts. 

Any two devices connected via a SRC protocol must reside within a certain distance from each other (i.e., the communication range). 
For example, let $d\in D$ be some IoT device that supports SRC protocol $p\in SRC$, and let $h\in H$ be some host that supports the same protocol.  
If $d$ is deployed in location $l$ and $h$ resides within the communication range of $l$, then $d$ may communicate with $h$ and vice~versa. 
\begin{definition}[Location Range]
\label{def:range}
We define the $range:L\cup\{\bot\}\rightarrow 2^{D\cup H}$ 
of a particular location as the set of hosts that may communicate with an IoT device deployed there. 
\end{definition}

\review{It is important to note that $range(l), l\in L$ is an estimation based on the radio specification of different IoT devices. 
The actual set of devices in range of IoT device deploy in location $l$ may vary depending on the power of the radio, obstacles, interference, etc. }

For the ease of discussion we ignore the protocol type and the specifications of the devices in Definition~\ref{def:range}. 
The definition of $range$ can be augmented with this additional information without modifications to the algorithms presented. 
Please note that a device can be in the range of several locations and that no devices are in the range of the non-location $\bot$ (i.e. $range(\bot)=\emptyset$). 

\subsection{Attack Graph Definition}
\label{sec:ag_def}
The potential locations of IoT devices and SRC protocols are integrated in the attack graph analysis methodology after the scanning stage and before attack graph modeling.   
For every possible deployment of IoT devices, that will be considered during the course of the optimization, we augment the connectivity map of \emph{devices} to include the hypothetical connections between any IoT device $d\in D$ deployed in location $depl(d)$ and all devices in the range of $d$ : $range(depl(d))$. 

Once the connectivity between all devices has been defined, we use the standard MulVAL framework to generate an attack graph that considers some given deployment of IoT devices. Each deployment has a different attack graph, depending on the devices deployed. If no IoT device is deployed the deployment is empty ($depl_{empty}$), and the attack graph is simply the original attack graph of the organization.

We adopt the attack graph definition introduced by Ou \etal~\citep{ou2006scalable}.
\begin{definition}[Logical Attack Graph]
\label{def:ag}
Let $depl$ be a deployment of IoT devices in an organization. 
The logical attack graph $G_{depl}$ is a tuple:
$$G_{depl}=(N_p, N_e, N_f, E, M, g),$$
where $N_p$, $N_e$, and $N_f$ are the sets of privilege nodes, exploit nodes and fact (leaf) nodes, respectively, and $E$ is a set of directed edges
$$E\subseteq (N_p\times N_e) \cup (N_e\times (N_e\cup N_f)),$$
\end{definition}

There are two types of edges in an attack graph. 
An edge $(e,p) \in E$  from an exploit node $e\in N_e$ to a privilege node $p\in N_p$ means that the attacker can gain privilege $p$ by executing exploit $e$. 
In order to gain a privilege, an attacker needs to execute one of the exploits leading to it.

An edge $(f,e) \in E$ from a fact node or a privilege node $f\in N_f\cup N_p$ to an exploit node $e\in N_e$ means that the node $f$ is a precondition to executing the exploit $e$. 
\review{For example, a fact node could be a vulnerability in the Bluetooth protocol that can be exploited if the attacker is in the Bluetooth range of the vulnerable device. }
In order to execute an exploit, the attacker needs all of the privileges and facts that lead to the exploit.

In this paper, in contrast to the definition introduced by Ou \etal~\citep{ou2006scalable}, the edge orientations follow the direction of the implied logical operation.

Next, we define the term \emph{attack plan}. For that purpose, we changed the notations from Gefen \etal~\citep{gefen2012pruning} slightly, as follows:

$pre(e) = \{ v \in N_p \cup N_f | (v,e) \in E \}$ are all of the preconditions of node $e$.

$obt(p) = \{ e \in N_e | v \in N_p \& (e,v) \in E \}$ is the set of exploits that lead to privilege node $p$ \review{(the set of privileges the attacker obtained)}.

An attack plan is a sub-graph $G_{depl}'$ of some attack graph $G_{depl}$ that represents a scenario in which the attacker manages to reach the goal, namely $g \in G_{depl}'$.
Therefore, in an attack plan all of the preconditions of an exploit $e\in G_{depl}'$ are satisfied, and each privilege $p\in G_{depl}'$ is obtained by an exploit.

\begin{definition}[Attack Plan]
Let $AP(G_{depl})$ be all of the attack plans of graph $G_{depl}$.
Each attack plan $G_{depl}'\in AP(G_{depl})$ needs to satisfy these three conditions:
\begin{itemize}
    \item $g \in G_{depl}'$
    \item $\forall a \in N_e : pre(a) \subseteq G_{depl}' | N_e \in G_{depl}' $
    \item $\forall p \in N_p : \exists a \in obt(p) \subseteq G_{depl}' | N_p \in G_{depl}' $
\end{itemize}

We consider the length of an attack plan as the number of nodes it contains.
$OptLen(G_{depl})$ is the length of the shortest attack plan in graph $G$, a $OptCnt(G_{depl})$ indicates how many of the shortest attack plans  there are in graph $G$.
\end{definition}


\subsection{Risk Score}
\label{sec:risk_score}
%

The network security can be estimated by the Risk Score, where the higher the risk score the lower the security of the network.
In an environment in which IoT devices are deployed, there are a few aspects to consider when choosing a method for computing the risk~score. 

First, the method needs to convey that the deployment of IoT devices may generate new attack plans.
Consequently, the cost of an attack may drop and the likelihood of an attack may increase due to the additional vulnerabilities and opportunities for lateral movement that an attacker can exploit.
Second, the method needs to indicate the changes in different deployments and be sensitive enough to detect the changes caused by the deployment of even a single additional IoT device.

We consider a deployment of IoT devices that reduces the number of options the attacker has for an attack. 
Therefore, in our work, we choose to calculate the shortest attack plans, taking their length and quantity into consideration. 
Gonda \etal~\citep{gonda2017ranking} describes the computation of the shortest attack plans in detail. 
As noted by the authors, enumerating all of the attack plans is NP-hard, which means that the running time can be exponential, however, we performed this computation on several networks, and the running time was short, as can also be seen in Section~\ref{sec:results}.

\begin{definition}[Risk Score]
$R(depl)$ is a tuple that represents the risk score of deployment $depl$.
The first element is the length of the shortest attack plan in graph $G_{depl}$, and the second element indicates how many of the shortest attack plans there are.
$$R(depl) = (OptLen(G_{depl})), OptCnt(G_{depl}))$$
\end{definition}

As mentioned above, we took two aspects of the shortest plans into consideration: the length of the plan and how many of the shortest plans there are.
For example, the risk score for the scenario in Example~\ref{run_example} is $R(15,1)$, since there is only one attack plan, and this plan has all fifteen  nodes in the graph (see Figure~\ref{fig:attack_graph}).

Considering only one of the above, the number of shortest plans or the length of the shortest plan, will not provide a good estimation of network security .
Suppose a network has $x$ shortest plans of length $l$ to the goal.
Further suppose that after deploying an IoT device, we now have a new plan of length $z$ that leads to the goal, when $z<l$.
In this case, the total number of shortest plans will decrease to one ($1<x$). If we only took into account how many of the shortest plans there are, it would appear that the risk score decreased (from $x$ to one), which implies that the network is now more secure. 
However, adding a device does not, in itself, eliminate any plans (i.e., all of the plans that existed before the device was added still exist). 
Therefore, adding a device can only create new plans, and the security risk can only increase.
Only considering the length of the shortest plan is also problematic, since a network with one plan of length $x$ is much more secure than a network with multiple plans of length $x$. 


For each comparison of the risk scores of various deployments, we compared the length of the shortest plans, and if the shortest plans in each deployment were equal, we considered the number of the shortest plans. 
Intuitively, the risk increases as the possible attack plans become shorter and as more of the shortest attack plans are added.
\begin{definition}[Deployment Comparison]
Let $depl_X$ and $depl_Y$ be two deployments of IoT devices. 
We say that  $depl_X$ is superior to $depl_Y$, denoted as $depl_X \prec depl_Y$, if and only if 
$$OptLen(G_{depl_x}) > OptLen(G_{depl_y}) \vee $$ 
$$[OptLen(G_{depl_x})=OptLen(G_{depl_y}) \wedge $$
$$ OptCnt(G_{depl_x})< OptCnt(G_{depl_y})]$$
\end{definition}


\section{Deployment Optimization Problem}
\label{sec:deployment_opt}
In this section, we introduce the terms and notation used to define the two IoT deployment optimization problems: (1) Full Deployment with Minimal Risk (FDMR), and (2) Maximal Utility without Risk Deterioration (MURD).

\textbf{FDMR Problem.} 
Given an attack graph of an organization $G$, a set of IoT devices $D$ of types $T$, and the location constraints $C$, 
find the deployment ($depl_{full}$) of IoT devices such that all of the IoT devices are deployed subject to location constraints, and the risk score $R(depl_{full})$ is minimized.  
\begin{definition}[Full Deployment with Minimal Risk (FDMR) Problem]
Given the four-tuple $<G,D,T,C>$, find $depl_{full}$ such that $R(depl_{full})$ is minimized 
$$\argmin_{depl_{full}}\{R(depl_{full})\}$$
\end{definition}

\textbf{MURD Problem.}
Given an attack graph of an organization $G$, a set of IoT devices $D$ of types $T$, and the location constraints $C$, find the deployment that consists of the highest number of IoT devices without increasing the risk score $R$. 
\begin{definition}[Maximal Utility without Risk Deterioration (MURD) Problem]
Given the four-tuple $<G,D,T,C>$, find $depl$ such that $|R(depl)|$ is maximized and $R(depl)$ = $R(depl_{empty})$ 
$$\argmax_{depl}\{|R(depl)| : R(depl) = R(depl_{empty})\}$$
\end{definition}

\subsection{\textbf{Search Space}}

Next, we define the search space for both FDMR and MURD. 
In each case, the state of the search space is organized as a binary tree where at each state a decision is made either to deploy (left child) or not to deploy (right child) a particular IoT device in a particular location. 
The root state is an empty deployment where no decisions have been made yet. 
Every path from the root node of the search space corresponds to a set of decisions. 
This means that a path from the root to any state defines where some of the IoT devices are deployed and where some other IoT devices cannot be deployed.
The set of left children along a path is a partial deployment of IoT devices. 
In this way, we consider all possible deployments, subject to location constraints.

For every node of the search space we derive the respective attack graph $G_{depl}$ and compute the risk score $R(depl)$. 
The goal nodes depend on the specific problem. 
In the FDMR problem the goal nodes include all states with a deployment that meets all of the constraints, $depl_{full}$ and the objective is to identify the goal state with the lowest risk score.
In the MURD problem the goal states include all states with a deployment that has the same risk score as the initial state.

\subsection{Search Algorithm}
\label{sec:heuristic}
For our heuristic search, we used the DFBnB algorithm (as described in Section~\ref{sec:back-search}). The heuristic function will described later in this section.
As we mentioned above, each state in our search tree has two children (left and right). 
In one, we added an IoT device to the deployment in a certain location, and in the other, we did not allow the IoT device to be deployed in that location. 
In practice, each state has various options regarding which IoT devices to deploy. 
We randomly chose one device ($d$) and one location ($l$) where $d$ can still be deployed and generate two children: deploy $d$ at $l$ and do not deploy $d$ at $l$.
For the left child corresponding to the deploy decision, we generate a new attack graph and recalculate the risk score and the value of the heuristic function.
We do not calculate the risk score for the right (do not deploy) child, as this child's risk score did not change, since the risk score depends only on the deployed devices.

\textbf{Heuristic Function.}
In order to calculate the heuristic functions, we created a table of risk scores $Table(depl_n)$ which contains the risk scores for each IoT device in each possible location. 
In other words, we simulate the deployment of a single IoT device each time. 
For each deployment, we update the table, removing the IoT device that was deployed or not allowed to be deployed. 
If the length of the shortest plan is shorter than the length of the shortest plan of the initial state, the heuristic's value in the table is zero.

\begin{definition}[FDMR Heuristic Function]
For the FDMR problem, the heuristic function underestimates the lowest possible change in risk in every subtree.
Then, whenever the risk score of the best full deployment found so far is lower than the risk score of any full deployment that can be found within a subtree, that subtree is pruned.

For FDMR, let $h_{FDMR}(depl_n)$ be the heuristic of $depl_n$.
$h_{FDMR}(depl_n)$ is the minimal $R(depl_d)$ 
and $R(depl_d) \in Table(depl_n)$.
$$ h_{FDMR}(depl_n) = \argmin_{d\in D}\{R(depl_d) \in Table(depl_n) \}$$
Intuitively, $h_{FDMR}$ underestimates the risk score because (1) individually each deployed device increases the risk according to $Table(depl_n)$, but (2) together multiple deployed devices may result in attack plans that were not accounted for yet.   
\end{definition}

\begin{definition}[MURD Heuristic Function]
For the MURD problem, the heuristic function overestimates the highest possible change in the number of IoT devices that can be deployed without increasing the risk. 
Then, whenever the number of devices deployed according to the incumbent solution found so far is larger than the number of devices that can possibly be deployed by continuing to search a subtree, that subtree is pruned.

We want to deploy the highest number of IoT devices possible, hence the heuristic function counts the number of IoT devices in $Table(depl_n)$ with the same risk score as the root state.
Let $h(depl_n)$ be the heuristic of $depl_n$. $h_{MURD}(depl_n)$ is the number of devices with a risk score equal to initial state $R(depl_{empty})$, such that 

$|R(depl_d)=R(depl_{empty})|$ and $R(depl_d) \in Table(depl_n)$.
$$ h_{MURD}(depl_n) = |R(depl_d) \in Table(depl_n) : R(depl_d)=R(depl_{empty})|$$
Intuitively, $h_{MURD}$ overestimates the number of devices that can be deployed because (1) any IoT device that increases the risk according to $Table(depl_n)$ cannot be deployed, and (2) even if individually a set of deployed devices does not increase the risk score, together they may result in an attack plan that was not available before.   
\end{definition}



\section{Evaluation}
\label{sec:evaluation}
We conducted experiment for each one of the problems we wish to solve: finding the full deployment with minimal risk (FDMR), and finding the maximal utility without risk deterioration (MURD).
For both problems, we used the suggested DFBnB algorithm with the heuristics described in Section~\ref{sec:heuristic}.

\subsection{Data Preparation}
\label{sec:data}
To evaluate our proposed method, we conducted a set of experiments using an attack graph that was derived from a real organization network.

\begin{figure}
\centering
\includegraphics[width=0.45\textwidth]
{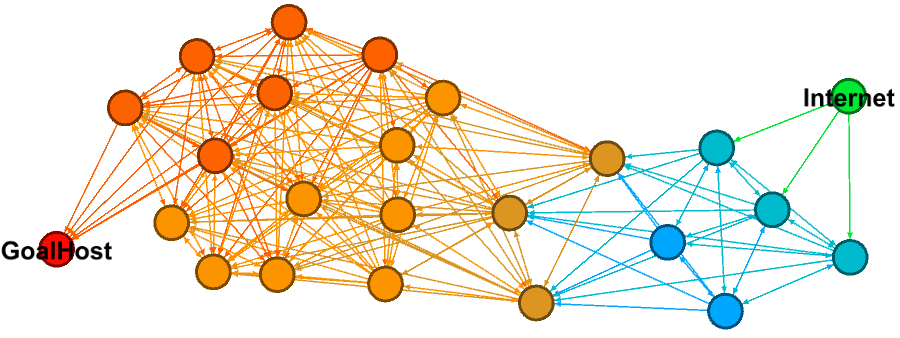}
\caption{Connectivity graph of the hosts in the organizational network, derived from the VLAN topology. } The different colors represent the different VLANs. \review{The blue nodes are DMZ VLAN, and the orange nodes are the internal organization network. Each node represents a host, and an edge indicates a connection between two hosts.} 
\label{fig:org_graph}
\end{figure}

\textbf{Organization Network.}
The network of the organization is a real network consisting of $24$ hosts which was used by Gonda \etal~\citep{gonda2017ranking}.
The network of the organization was scanned using Nessus Scanner, and then MulVAL was used to generate the attack graph based on the scanning results.
Figure~\ref{fig:org_graph} depicts the connectivity of the hosts in the network, \review{derived from the VLAN topology}. Each node represents a host, and an edge indicates a connection between two hosts.

An organization can have more than one host that it wishes to protect, and this is translated to multiple targets for the attacker.
To simplify things, all target hosts are connected to an abstract $goalHost$, and the goal of the attack graph is to execute code in this host.
Executing code on the $goalHost$ proves that the attacker managed to control one of the targeted hosts that led to the goal.
As part of the experimental setup we assume that the organization is free from inside adversaries and that the potential attacker is located on the $Internet$.
The attack graph has a host that represents the Internet.
Detailed information on the scanning process is provided in~\citep{gonda2017ranking}.

\textbf{Simulating IoT Devices.}
The network of the organization used in the experiment does not include any IoT devices.
Therefore, we opt to simulate the IoT devices, their communication protocols, and the constraints required for their deployment.
We simulated three IoT types (detector, refrigerator, camera), nine different IoT devices (four detectors, two cameras, and three refrigerators), and eight locations for the deployment of IoT devices.

In the simulation, the organization would like to deploy three detectors for which there are four possible locations, one camera for which there are two possible locations, and two refrigerators for which there are two possible locations. 
Therefore, a total of six IoT devices needed to be deployed.

Formally, as defined in Definition~\ref{def:locations_const}, the location constraints in our simulation are defined as follows:

$C(detector)$ = $(\{l_{det_1},l_{det_2},l_{det_3}, l_{det_4}\}$,$3$,$\{d_{det1},d_{det2},d_{det3}$\linebreak,$d_{det4}\})$

$C(camera) = (\{l_{cam_1},l_{cam_2}\},1,\{d_{cam1},d_{cam2}\})$

$C(refrigerator) = (\{l_{ref_1},l_{ref_2}\},2,\{d_{ref1},d_{ref2},d_{ref3}\})$

\review{
Using permutation\footnote{\review{Permutation $P^n_k$ mean that for $n$ items, we want to find the number of ways $k$ items can be ordered.}}
and combination,\footnote{\review{Combination $C^n_k$ is a selection of $k$ items from a collection of size $n$, such that the order of selection does not matter.}} we can calculate the total number of options in the search space.}

\review{
$C^4_3 P^4_3 \cdot C^2_1 P^2_1 \cdot C^2_2 P^3_2 = 2304$
}

Meaning, there are $2304$ possible deployments.


\textbf{Simulating Short-Range Communication.} 
We simulated two short-range communication protocols (ZigBee and Bluetooth) and randomly divided them between all IoT devices and hosts \review{so that 75\% of the hosts have Bluetooth and 20\% number of them have Zigbee, and 40\% of the IoT devices have Bluetooth and 90\% of them have Zigbee. 
}

\textbf{Simulating Vulnerabilities.} In order to create potential attack plans that include IoT devices, we simulated existing vulnerabilities that can be exploited as follows. \review{For each IoT device and for each host, in addition to its known vulnerabilities (from the scanning performed), we created a vulnerability based on the protocol used. }

\textbf{Simulating Physical Location of Hosts.} 
The actual physical location of the real hosts was unavailable. 
The location of the hosts is important in order to simulate the proximity of the IoT devices to the host, and consequently create potential attack plans involving the IoT devices. 
Therefore, we randomly divided the hosts among the eight simulated location ranges. 
Note that a host can be in proximity to more than one IoT device.

\subsection{Experimental Setup}
\review{The experiments were conducted on Hyper-V VM, with four virtual CPUs (two cores) and 8GB RAM. The setup of the experiments is as follow: }

\textbf{Number of Executions.} In order to strengthen the validity of our results, we executed the experiment forty times, using a different host location each time. 
In other words, we simulated the physical location of hosts forty times. 
The results in the next section are the average results of all executions.


\textbf{Evaluation Measures.} 
We computed two measures: the first is the execution time, and the second is the risk score of a suggested IoT deployment (for the FDMR use case) or the number of deployable IoT devices (for the MURD use case). 
The evaluation measures were averaged over the all of the executions. 
The execution time is important, since this can be a weak point, as one of the difficulties in attack graphs and solutions that are based on attack graphs is execution time.
    
\textbf{Random Deployment.} 
For comparison, we also ran both problems randomly as a baseline. 
This scenario represents an organization that randomly deploys IoT devices, without considering the security aspect. 
That is to say, for the FDMR problem we randomly deployed all IoT devices five times and took the average risk score of all the deployments.
In the MURD problem, each time we added a device randomly and computed the risk score. 
We started with no IoT devices deployed and continued until full deployment. We ran five times each number of devices. 
This random baseline was executed the same number of times as our algorithm (forty times).  

\subsection{Results}
\label{sec:results}

Table~\ref{tab:results} presents the results.
Note that the risk score only includes the  number of the shortest paths ($OptCnt(G_{depl})$). 
The length of the shortest paths in all of the results presented is $29$.

\begin{table*}
\caption{Results (average over 40 executions)}
\label{tab:results}
\begin{tabular}{@{}|c|c|c|c|c|c|@{}}
\toprule
\multirow{2}{*}{Problem} & \multicolumn{3}{c|}{DFBnB}                             & \multicolumn{2}{c|}{Random Deployment}    \\ \cmidrule(l){2-6} 
                         & Risk Score (std) & Devices Deployed (std) & Time (min) & Risk Score (std) & Devices Deployed (std) \\ \midrule
FDMR                     & 1229 (239.41)    & 6 (0)                  & 36.20      & 1494.46 (370.82) & 6 (0)                  \\ \midrule
MURD                     & 1032 (0)         & 4.40 (1.13)            & 3.88       & 1538.95 (364.74) & 4 (0)                  \\ \bottomrule
\end{tabular}
\end{table*}

\textbf{Full Deployment with Minimal Risk (FDMR).}
Full Deployment with Minimal Risk(FDMR). 
In the FDMR problem, the average risk score of all runs is $1229$, an increase of 19\% compared to the risk score without any IoT devices which is $1032$. 
The algorithm took an average of $36$ minutes to run, which is a reasonable amount of time and provides an indication of its feasibility on a larger scale.

\textbf{Maximal Utility without Risk Deterioration (MURD).}
In the MURD problem, the average number of IoT devices that can be deployed without affecting the security risk is $4.40$. This number means that, on average, four to five devices can be deployed without any change in the risk score.
It took the algorithm an average of $3.88$ minutes to compute, which is also a reasonable time.

\textbf{Random Deployment.}
In FDMR, the average risk score was $1494$, which is an increase of 44\% from the initial state. We can see that randomly deploying IoT devices leads to less safe network, compared to the increase of only 19\% when using our algorithm.

\begin{figure}
\centering
\includegraphics[width=0.45\textwidth]
{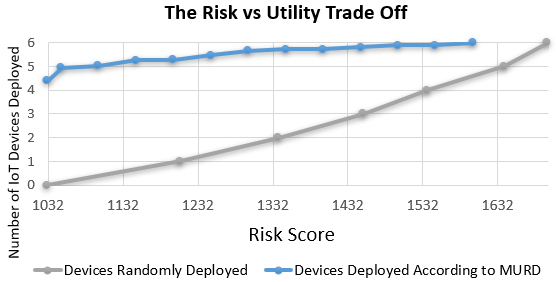}
\caption{The blue graph indicates the average number of devices deployed under security risk bound. The grey graph indicates the average risk score of deployments with each number of devices. 
} 
\label{fig:murd_stats}
\end{figure}

In the MURD problem, the average risk score of deploying four IoT devices is $1538$. We chose four devices because with our algorithm we managed to deploy an average of $4.40$ devices without influencing the security of the network. This result is also much higher than the basic risk score of $1032$, with no IoT devices deployed.
The average risk score of other numbers of devices can be seen in Figure ~\ref{fig:murd_stats}\review{~(in grey)}, where we present the average risk score of deployments with each number of devices, ranging from zero (empty deployment) to six (full deployment).

\textbf{Running Time.}
The average time for the algorithm to solve the FDMR problem was $36$ minutes, and for the MURD problem less than four minutes. In addition, the average time it took to compute the risk score in all of the executions on both problems was less than a second ($0.95$ seconds), and the average time to calculate the heuristic was $2.85e^{-5}$ seconds. 
It took $23$ seconds, on average, to compute the heuristic table before the start of the algorithm.
These measurements are very low and practical, suggesting  that the algorithm can run on additional networks. 

\textbf{Additional Results.}
\review{We investigated the trade-off between the allowed risk of the IoT deployment and the maximal number of IoT devices that can be deployed. 
Figure~\ref{fig:murd_stats} further emphasizes the difference between random and optimal deployment of IoT devices.  
On one hand, 4-5 randomly deployed IoT devices increase the number of possible attack plans by ~50\%. 
On the other hand the same number of IoT devices can be deployed with insignificant risk deterioration.      
We can also see from Figure~\ref{fig:murd_stats} that the difference between optimal and random deployment strategies diminishes as we try to deploy six IoT devices. 
}

\begin{figure}
\centering
\includegraphics[width=0.45\textwidth]
{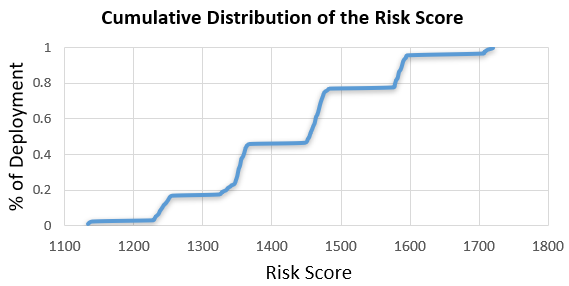}
\caption{The $x$-axis is the cumulative risk score, and the $y$-axis is the percentage of deployments for which the risk score is less than $x$.
} 
\label{fig:dist}
\end{figure}

Figure~\ref{fig:dist} illustrates the challenge in finding the safest deployment of IoT devices.
The graph presents the cumulative distribution of the risk scores of all deployments in one execution. 
The $x$-axis is the cumulative risk score, and the $y$-axis is the percentage of deployments for which the risk score is less than $x$. 
As can be seen, 50\% of the deployments have a risk score lower than $1458$. 
Moreover, only $16$ deployments (0.7\% of all deployments) are optimal, with a risk score of $1134$, i.e., the chances of a random selection to choose an optimal deployment in that execution was~$0.007$.

\review{The risk score of an optimal deployment may change when new vulnerabilities are discovered, leading to potentially inferior deployment. 
To conclude the experimental evaluation we tested the robustness of the optimal deployment 
of an arbitrary execution from the FDMR problem, with risk score of $1134$. 
We perturbed vulnerabilities of 10\% and 20\% of the devices in the network by discarding all current vulnerabilities of the chosen devices and randomly assigning new vulnerabilities as described in Section~\ref{sec:data}. 
This process was repeated 10 times. 
The average risk score of the optimal deployment after changing 10\% and 20\% of the vulnerabilities varied by 1\%-10\% in both directions. 
Some times the risk of the optimal deployment increased and some times it dropped. 
Overall the changes in the risk of the optimal deployment due to perturbation of the vulnerabilities were not statistically significant. 
}



\section{Conclusion and Future Work}
\label{sec:conculsion}
We present a novel method for suggesting the optimal deployment (in terms of the security risk) of a set of IoT devices within an organization. 
In order to accomplish this, we augmented the conventional attack graph to include short-range communication protocols inherent to IoT devices.
To the best of our knowledge, this is the first work that takes the physical location of devices and different communication protocols into account.


We demonstrated the importance of planning a deployment of IoT devices by solving two scenarios, approaching them as an optimization problem. 
We proposed a novel method for evaluating the risk of IoT device deployment using an augmented attack graph, and used the proposed method to address these two scenarios. 
Our results revealed the potential risk in deploying IoT devices in organizations and showed that randomly deploying devices can greatly affect the security of the organization's network. 
We solved the two scenarios on a real organization with a small to medium sized network, with a running time of less than an hour.

\review{Our algorithm, and in particular, our heuristic approach, assumes that the potential risk of two deployed devices is greater than or equal to the sum of their individual risk scores. Any method of risk calculation that satisfies this assumption can be used in the algorithm. }
\review{The method of risk score calculation used in this paper has some limitations. It does not take the cost of different exploits into account, which can be a major consideration for an attacker. As a result, the method does not capture the heterogeneity and homogeneity of vulnerabilities along an attack path. In addition, the method only considers the shortest paths, but an attacker can choose a longer path, for various reasons. }

\review{Future work may extend the current research in the following directions.
First, it is desirable to increase sizes of the attack graph that can be optimized by providing more accurate heuristic functions. 
In addition, the optimization methods proposed in this paper should be tested with variety of risk scores that encompass the true cost of the attack, the probability of the attack success, or both. 
Finally, cyber-physical capabilities of IoT devices as well as their unique functionalities should be incorporated into an extended model. }

\bibliographystyle{ACM-Reference-Format}
\bibliography{bibliography}


\begin{thebibliography}{49}


\ifx \showCODEN    \undefined \def \showCODEN     #1{\unskip}     \fi
\ifx \showDOI      \undefined \def \showDOI       #1{#1}\fi
\ifx \showISBNx    \undefined \def \showISBNx     #1{\unskip}     \fi
\ifx \showISBNxiii \undefined \def \showISBNxiii  #1{\unskip}     \fi
\ifx \showISSN     \undefined \def \showISSN      #1{\unskip}     \fi
\ifx \showLCCN     \undefined \def \showLCCN      #1{\unskip}     \fi
\ifx \shownote     \undefined \def \shownote      #1{#1}          \fi
\ifx \showarticletitle \undefined \def \showarticletitle #1{#1}   \fi
\ifx \showURL      \undefined \def \showURL       {\relax}        \fi
\providecommand\bibfield[2]{#2}
\providecommand\bibinfo[2]{#2}
\providecommand\natexlab[1]{#1}
\providecommand\showeprint[2][]{arXiv:#2}

\bibitem[\protect\citeauthoryear{Abadi and Jalili}{Abadi and Jalili}{2006}]%
        {abadi2006ant}
\bibfield{author}{\bibinfo{person}{M Abadi} {and} \bibinfo{person}{S Jalili}.}
  \bibinfo{year}{2006}\natexlab{}.
\newblock \showarticletitle{An ant colony optimization algorithm for network
  vulnerability analysis}.
\newblock \bibinfo{journal}{\emph{Iranian Journal of Electrical and Electronic
  Engineering}} \bibinfo{volume}{2}, \bibinfo{number}{3}
  (\bibinfo{year}{2006}), \bibinfo{pages}{106--120}.
\newblock


\bibitem[\protect\citeauthoryear{Abie and Balasingham}{Abie and
  Balasingham}{2012}]%
        {abie2012risk}
\bibfield{author}{\bibinfo{person}{Habtamu Abie} {and} \bibinfo{person}{Ilangko
  Balasingham}.} \bibinfo{year}{2012}\natexlab{}.
\newblock \showarticletitle{Risk-based adaptive security for smart IoT in
  eHealth}. In \bibinfo{booktitle}{\emph{Proceedings of the 7th International
  Conference on Body Area Networks}}. ICST (Institute for Computer Sciences,
  Social-Informatics and~…, \bibinfo{pages}{269--275}.
\newblock


\bibitem[\protect\citeauthoryear{Almohri, Watson, Yao, and Ou}{Almohri
  et~al\mbox{.}}{2016}]%
        {almohri2016security}
\bibfield{author}{\bibinfo{person}{Hussain~MJ Almohri},
  \bibinfo{person}{Layne~T Watson}, \bibinfo{person}{Danfeng Yao}, {and}
  \bibinfo{person}{Xinming Ou}.} \bibinfo{year}{2016}\natexlab{}.
\newblock \showarticletitle{Security optimization of dynamic networks with
  probabilistic graph modeling and linear programming}.
\newblock \bibinfo{journal}{\emph{IEEE Transactions on Dependable and Secure
  Computing}} \bibinfo{volume}{13}, \bibinfo{number}{4} (\bibinfo{year}{2016}),
  \bibinfo{pages}{474--487}.
\newblock


\bibitem[\protect\citeauthoryear{Ammann, Wijesekera, and Kaushik}{Ammann
  et~al\mbox{.}}{2002}]%
        {Ammann2002ScalableAnalysis}
\bibfield{author}{\bibinfo{person}{Paul Ammann}, \bibinfo{person}{Duminda
  Wijesekera}, {and} \bibinfo{person}{Saket Kaushik}.}
  \bibinfo{year}{2002}\natexlab{}.
\newblock \showarticletitle{{Scalable, graph-based network vulnerability
  analysis}}.
\newblock \bibinfo{journal}{\emph{Proceedings of the 9th ACM conference on
  Computer and communications security - CCS '02}} (\bibinfo{year}{2002}),
  \bibinfo{pages}{217}.
\newblock
\showISBNx{1581133855}
\urldef\tempurl%
\url{https://doi.org/10.1145/586110.586140}
\showDOI{\tempurl}


\bibitem[\protect\citeauthoryear{Bandyopadhyay and Sen}{Bandyopadhyay and
  Sen}{2011}]%
        {Bandyopadhyay2011InternetStandardization}
\bibfield{author}{\bibinfo{person}{Debasis Bandyopadhyay} {and}
  \bibinfo{person}{Jaydip Sen}.} \bibinfo{year}{2011}\natexlab{}.
\newblock \showarticletitle{{Internet of things: Applications and challenges in
  technology and standardization}}.
\newblock \bibinfo{journal}{\emph{Wireless Personal Communications}}
  \bibinfo{volume}{58}, \bibinfo{number}{1} (\bibinfo{year}{2011}),
  \bibinfo{pages}{49--69}.
\newblock
\showISBNx{0929-6212}
\showISSN{09296212}
\urldef\tempurl%
\url{https://doi.org/10.1007/s11277-011-0288-5}
\showDOI{\tempurl}


\bibitem[\protect\citeauthoryear{Barcena and Wueest}{Barcena and
  Wueest}{2015}]%
        {InsecuritySymantec}
\bibfield{author}{\bibinfo{person}{Mario~Ballano Barcena} {and}
  \bibinfo{person}{Candid Wueest}.} \bibinfo{year}{2015}\natexlab{}.
\newblock \bibinfo{booktitle}{\emph{{Insecurity in the Internet of Things}}}.
\newblock \bibinfo{type}{{T}echnical {R}eport} March.
\newblock


\bibitem[\protect\citeauthoryear{Beale, Deraison, Meer, Temmingh, and
  Walt}{Beale et~al\mbox{.}}{2004}]%
        {beale2004nessus}
\bibfield{author}{\bibinfo{person}{Jay Beale}, \bibinfo{person}{Renaud
  Deraison}, \bibinfo{person}{Haroon Meer}, \bibinfo{person}{Roelof Temmingh},
  {and} \bibinfo{person}{Charl Van~Der Walt}.} \bibinfo{year}{2004}\natexlab{}.
\newblock \bibinfo{booktitle}{\emph{Nessus network auditing}}.
\newblock \bibinfo{publisher}{Syngress Publishing}.
\newblock


\bibitem[\protect\citeauthoryear{Chen, Liu, Zhang, and Su}{Chen
  et~al\mbox{.}}{2010}]%
        {chen2010scalable}
\bibfield{author}{\bibinfo{person}{Feng Chen}, \bibinfo{person}{Dehui Liu},
  \bibinfo{person}{Yi Zhang}, {and} \bibinfo{person}{Jinshu Su}.}
  \bibinfo{year}{2010}\natexlab{}.
\newblock \showarticletitle{A scalable approach to analyzing network security
  using compact attack graphs}.
\newblock \bibinfo{journal}{\emph{Journal of Networks}} \bibinfo{volume}{5},
  \bibinfo{number}{5} (\bibinfo{year}{2010}), \bibinfo{pages}{543}.
\newblock


\bibitem[\protect\citeauthoryear{Cope, Campbell, and Hayajneh}{Cope
  et~al\mbox{.}}{2017}]%
        {cope2017investigation}
\bibfield{author}{\bibinfo{person}{Peter Cope}, \bibinfo{person}{Joseph
  Campbell}, {and} \bibinfo{person}{Thaier Hayajneh}.}
  \bibinfo{year}{2017}\natexlab{}.
\newblock \showarticletitle{An investigation of Bluetooth security
  vulnerabilities}. In \bibinfo{booktitle}{\emph{2017 IEEE 7th Annual Computing
  and Communication Workshop and Conference (CCWC)}}. IEEE,
  \bibinfo{pages}{1--7}.
\newblock


\bibitem[\protect\citeauthoryear{Dunning}{Dunning}{2010}]%
        {dunning2010taming}
\bibfield{author}{\bibinfo{person}{John Dunning}.}
  \bibinfo{year}{2010}\natexlab{}.
\newblock \showarticletitle{Taming the blue beast: A survey of bluetooth based
  threats}.
\newblock \bibinfo{journal}{\emph{IEEE Security \& Privacy}}
  \bibinfo{volume}{8}, \bibinfo{number}{2} (\bibinfo{year}{2010}),
  \bibinfo{pages}{20--27}.
\newblock


\bibitem[\protect\citeauthoryear{Ge, Hong, Alzaid, and Kim}{Ge
  et~al\mbox{.}}{2017}]%
        {Ge2017SecurityDevices}
\bibfield{author}{\bibinfo{person}{Mengmeng Ge}, \bibinfo{person}{Jin~B. Hong},
  \bibinfo{person}{Hani Alzaid}, {and} \bibinfo{person}{Dong~Seong Kim}.}
  \bibinfo{year}{2017}\natexlab{}.
\newblock \showarticletitle{{Security modeling and analysis of cross-protocol
  IoT devices}}.
\newblock \bibinfo{journal}{\emph{Proceedings - 16th IEEE International
  Conference on Trust, Security and Privacy in Computing and Communications,
  11th IEEE International Conference on Big Data Science and Engineering and
  14th IEEE International Conference on Embedded Software and Systems}}
  (\bibinfo{year}{2017}), \bibinfo{pages}{1043--1048}.
\newblock
\showISBNx{9781509049059}
\urldef\tempurl%
\url{https://doi.org/10.1109/Trustcom/BigDataSE/ICESS.2017.350}
\showDOI{\tempurl}


\bibitem[\protect\citeauthoryear{Ge and Kim}{Ge and Kim}{2016}]%
        {Ge2016}
\bibfield{author}{\bibinfo{person}{Mengmeng Ge} {and}
  \bibinfo{person}{Dong~Seong Kim}.} \bibinfo{year}{2016}\natexlab{}.
\newblock \showarticletitle{{A framework for modeling and assessing security of
  the internet of things}}. In \bibinfo{booktitle}{\emph{Proceedings of the
  International Conference on Parallel and Distributed Systems - ICPADS}}.
\newblock
\showISBNx{9780769557854}
\showISSN{15219097}
\urldef\tempurl%
\url{https://doi.org/10.1109/ICPADS.2015.102}
\showDOI{\tempurl}


\bibitem[\protect\citeauthoryear{Gefen and Brafman}{Gefen and Brafman}{2012}]%
        {gefen2012pruning}
\bibfield{author}{\bibinfo{person}{Avitan Gefen} {and} \bibinfo{person}{Ronen
  Brafman}.} \bibinfo{year}{2012}\natexlab{}.
\newblock \showarticletitle{Pruning methods for optimal delete-free planning}.
  In \bibinfo{booktitle}{\emph{Twenty-Second International Conference on
  Automated Planning and Scheduling}}.
\newblock


\bibitem[\protect\citeauthoryear{Gonda, Shani, Puzis, and Shapira}{Gonda
  et~al\mbox{.}}{2017}]%
        {gonda2017ranking}
\bibfield{author}{\bibinfo{person}{Tom Gonda}, \bibinfo{person}{Guy Shani},
  \bibinfo{person}{Rami Puzis}, {and} \bibinfo{person}{Bracha Shapira}.}
  \bibinfo{year}{2017}\natexlab{}.
\newblock \showarticletitle{Ranking vulnerability fixes using planning graph
  analysis}. In \bibinfo{booktitle}{\emph{IWAISe: First International Workshop
  on Artificial Intelligence in Security}}. \bibinfo{pages}{41}.
\newblock


\bibitem[\protect\citeauthoryear{Hong and Kim}{Hong and Kim}{2012}]%
        {hong2012harms}
\bibfield{author}{\bibinfo{person}{Jin Hong} {and} \bibinfo{person}{Dong-Seong
  Kim}.} \bibinfo{year}{2012}\natexlab{}.
\newblock \showarticletitle{Harms: Hierarchical attack representation models
  for network security analysis}.
\newblock  (\bibinfo{year}{2012}).
\newblock


\bibitem[\protect\citeauthoryear{Huang, Meng, Gong, Liu, and Duan}{Huang
  et~al\mbox{.}}{2014}]%
        {huang2014novel}
\bibfield{author}{\bibinfo{person}{Jun Huang}, \bibinfo{person}{Yu Meng},
  \bibinfo{person}{Xuehong Gong}, \bibinfo{person}{Yanbing Liu}, {and}
  \bibinfo{person}{Qiang Duan}.} \bibinfo{year}{2014}\natexlab{}.
\newblock \showarticletitle{A novel deployment scheme for green internet of
  things}.
\newblock \bibinfo{journal}{\emph{IEEE Internet of Things Journal}}
  \bibinfo{volume}{1}, \bibinfo{number}{2} (\bibinfo{year}{2014}),
  \bibinfo{pages}{196--205}.
\newblock


\bibitem[\protect\citeauthoryear{Islam and Wang}{Islam and Wang}{2008}]%
        {islam2008heuristic}
\bibfield{author}{\bibinfo{person}{Tania Islam} {and} \bibinfo{person}{Lingyu
  Wang}.} \bibinfo{year}{2008}\natexlab{}.
\newblock \showarticletitle{A heuristic approach to minimum-cost network
  hardening using attack graph}. In \bibinfo{booktitle}{\emph{2008 New
  Technologies, Mobility and Security}}. IEEE, \bibinfo{pages}{1--5}.
\newblock


\bibitem[\protect\citeauthoryear{Jun-chun, Yong-jun, Ji-yin, and Shan}{Jun-chun
  et~al\mbox{.}}{2011}]%
        {jun2011minimum}
\bibfield{author}{\bibinfo{person}{MA Jun-chun}, \bibinfo{person}{WANG
  Yong-jun}, \bibinfo{person}{SUN Ji-yin}, {and} \bibinfo{person}{CHEN Shan}.}
  \bibinfo{year}{2011}\natexlab{}.
\newblock \showarticletitle{A minimum cost of network hardening model based on
  attack graphs}.
\newblock \bibinfo{journal}{\emph{Procedia Engineering}}  \bibinfo{volume}{15}
  (\bibinfo{year}{2011}), \bibinfo{pages}{3227--3233}.
\newblock


\bibitem[\protect\citeauthoryear{Korf}{Korf}{2010}]%
        {Korf2010ArtificialAlgorithms}
\bibfield{author}{\bibinfo{person}{Richard~E. Korf}.}
  \bibinfo{year}{2010}\natexlab{}.
\newblock \showarticletitle{{Artificial Intelligence Search Algorithms}}.
\newblock In \bibinfo{booktitle}{\emph{Algorithms and theory of computation
  handbook}}, \bibfield{editor}{\bibinfo{person}{Mikhail~J. Atallah} {and}
  \bibinfo{person}{Marina Blanton}} (Eds.). \bibinfo{pages}{22}.
\newblock
\showISBNx{978-1-58488-820-8}


\bibitem[\protect\citeauthoryear{Liu, Zhang, Zeng, Peng, and Chen}{Liu
  et~al\mbox{.}}{2012}]%
        {liu2012research}
\bibfield{author}{\bibinfo{person}{Caiming Liu}, \bibinfo{person}{Yan Zhang},
  \bibinfo{person}{Jinquan Zeng}, \bibinfo{person}{Lingxi Peng}, {and}
  \bibinfo{person}{Run Chen}.} \bibinfo{year}{2012}\natexlab{}.
\newblock \showarticletitle{Research on Dynamical Security Risk Assessment for
  the Internet of Things inspired by immunology}. In
  \bibinfo{booktitle}{\emph{Natural Computation (ICNC), 2012 Eighth
  International Conference on}}. IEEE, \bibinfo{pages}{874--878}.
\newblock


\bibitem[\protect\citeauthoryear{Meulen}{Meulen}{2017}]%
        {Meulen2017Gartner2016}
\bibfield{author}{\bibinfo{person}{Rob van~der Meulen}.}
  \bibinfo{year}{2017}\natexlab{}.
\newblock \showarticletitle{{Gartner Says 8.4 Billion Connected "Things" Will
  Be in Use in 2017, Up 31 Percent From 2016}}.
\newblock \bibinfo{journal}{\emph{Gartner}} (\bibinfo{year}{2017}),
  \bibinfo{pages}{1}.
\newblock
\urldef\tempurl%
\url{https://www.gartner.com/newsroom/id/3598917}
\showURL{%
\tempurl}


\bibitem[\protect\citeauthoryear{Minar and Tarique}{Minar and Tarique}{2012}]%
        {minar2012bluetooth}
\bibfield{author}{\bibinfo{person}{Nateq Be-Nazir~Ibn Minar} {and}
  \bibinfo{person}{Mohammed Tarique}.} \bibinfo{year}{2012}\natexlab{}.
\newblock \showarticletitle{Bluetooth security threats and solutions: a
  survey}.
\newblock \bibinfo{journal}{\emph{International Journal of Distributed and
  Parallel Systems}} \bibinfo{volume}{3}, \bibinfo{number}{1}
  (\bibinfo{year}{2012}), \bibinfo{pages}{127}.
\newblock


\bibitem[\protect\citeauthoryear{Mohsin, Sardar, Hasan, and Anwar}{Mohsin
  et~al\mbox{.}}{2017}]%
        {mohsin2017iotriskanalyzer}
\bibfield{author}{\bibinfo{person}{Mujahid Mohsin},
  \bibinfo{person}{Muhammad~Usama Sardar}, \bibinfo{person}{Osman Hasan}, {and}
  \bibinfo{person}{Zahid Anwar}.} \bibinfo{year}{2017}\natexlab{}.
\newblock \showarticletitle{IoTRiskAnalyzer: A Probabilistic Model Checking
  Based Framework for Formal Risk Analytics of the Internet of Things.}
\newblock \bibinfo{journal}{\emph{IEEE Access}}  \bibinfo{volume}{5}
  (\bibinfo{year}{2017}), \bibinfo{pages}{5494--5505}.
\newblock


\bibitem[\protect\citeauthoryear{Morgner, Mattejat, and Benenson}{Morgner
  et~al\mbox{.}}{2016}]%
        {morgner2016all}
\bibfield{author}{\bibinfo{person}{Philipp Morgner}, \bibinfo{person}{Stephan
  Mattejat}, {and} \bibinfo{person}{Zinaida Benenson}.}
  \bibinfo{year}{2016}\natexlab{}.
\newblock \showarticletitle{All your bulbs are belong to us: Investigating the
  current state of security in connected lighting systems}.
\newblock \bibinfo{journal}{\emph{arXiv preprint arXiv:1608.03732}}
  (\bibinfo{year}{2016}).
\newblock


\bibitem[\protect\citeauthoryear{Noel and Jajodia}{Noel and Jajodia}{2008}]%
        {noel2008optimal}
\bibfield{author}{\bibinfo{person}{Steven Noel} {and} \bibinfo{person}{Sushil
  Jajodia}.} \bibinfo{year}{2008}\natexlab{}.
\newblock \showarticletitle{Optimal ids sensor placement and alert
  prioritization using attack graphs}.
\newblock \bibinfo{journal}{\emph{Journal of Network and Systems Management}}
  \bibinfo{volume}{16}, \bibinfo{number}{3} (\bibinfo{year}{2008}),
  \bibinfo{pages}{259--275}.
\newblock


\bibitem[\protect\citeauthoryear{Noel and Jajodia}{Noel and Jajodia}{2014}]%
        {noel2014metrics}
\bibfield{author}{\bibinfo{person}{Steven Noel} {and} \bibinfo{person}{Sushil
  Jajodia}.} \bibinfo{year}{2014}\natexlab{}.
\newblock \showarticletitle{Metrics suite for network attack graph analytics}.
  In \bibinfo{booktitle}{\emph{Proceedings of the 9th Annual Cyber and
  Information Security Research Conference}}. ACM, \bibinfo{pages}{5--8}.
\newblock


\bibitem[\protect\citeauthoryear{Noel, Jajodia, O'Berry, and Jacobs}{Noel
  et~al\mbox{.}}{2003}]%
        {noel2003efficient}
\bibfield{author}{\bibinfo{person}{Steven Noel}, \bibinfo{person}{Sushil
  Jajodia}, \bibinfo{person}{Brian O'Berry}, {and} \bibinfo{person}{Michael
  Jacobs}.} \bibinfo{year}{2003}\natexlab{}.
\newblock \showarticletitle{Efficient minimum-cost network hardening via
  exploit dependency graphs}. In \bibinfo{booktitle}{\emph{19th Annual Computer
  Security Applications Conference, 2003. Proceedings.}} IEEE,
  \bibinfo{pages}{86--95}.
\newblock


\bibitem[\protect\citeauthoryear{Ou, Boyer, and McQueen}{Ou
  et~al\mbox{.}}{2006}]%
        {ou2006scalable}
\bibfield{author}{\bibinfo{person}{Xinming Ou}, \bibinfo{person}{Wayne~F
  Boyer}, {and} \bibinfo{person}{Miles~A McQueen}.}
  \bibinfo{year}{2006}\natexlab{}.
\newblock \showarticletitle{A scalable approach to attack graph generation}. In
  \bibinfo{booktitle}{\emph{Proceedings of the 13th ACM conference on Computer
  and communications security}}. ACM, \bibinfo{pages}{336--345}.
\newblock


\bibitem[\protect\citeauthoryear{Ou and Govindavajhala}{Ou and
  Govindavajhala}{2005}]%
        {ou2005mulval}
\bibfield{author}{\bibinfo{person}{Xinming Ou} {and} \bibinfo{person}{Sudhakar
  Govindavajhala}.} \bibinfo{year}{2005}\natexlab{}.
\newblock \showarticletitle{Mulval: A logic-based network security analyzer}.
  In \bibinfo{booktitle}{\emph{In 14th USENIX Security Symposium}}. Citeseer.
\newblock


\bibitem[\protect\citeauthoryear{Phillips and Swiler}{Phillips and
  Swiler}{1998}]%
        {phillips1998graph}
\bibfield{author}{\bibinfo{person}{Cynthia Phillips} {and}
  \bibinfo{person}{Laura~Painton Swiler}.} \bibinfo{year}{1998}\natexlab{}.
\newblock \showarticletitle{A graph-based system for network-vulnerability
  analysis}. In \bibinfo{booktitle}{\emph{Proceedings of the 1998 workshop on
  New security paradigms}}. ACM, \bibinfo{pages}{71--79}.
\newblock


\bibitem[\protect\citeauthoryear{Polad, Puzis, and Shapira}{Polad
  et~al\mbox{.}}{2017}]%
        {polad2017attack}
\bibfield{author}{\bibinfo{person}{Hadar Polad}, \bibinfo{person}{Rami Puzis},
  {and} \bibinfo{person}{Bracha Shapira}.} \bibinfo{year}{2017}\natexlab{}.
\newblock \showarticletitle{Attack graph obfuscation}. In
  \bibinfo{booktitle}{\emph{International Conference on Cyber Security
  Cryptography and Machine Learning}}. Springer, \bibinfo{pages}{269--287}.
\newblock


\bibitem[\protect\citeauthoryear{Roman, Najera, and Lopez}{Roman
  et~al\mbox{.}}{2011}]%
        {roman2011securing}
\bibfield{author}{\bibinfo{person}{Rodrigo Roman}, \bibinfo{person}{Pablo
  Najera}, {and} \bibinfo{person}{Javier Lopez}.}
  \bibinfo{year}{2011}\natexlab{}.
\newblock \showarticletitle{Securing the internet of things}.
\newblock \bibinfo{journal}{\emph{Computer}} \bibinfo{volume}{44},
  \bibinfo{number}{9} (\bibinfo{year}{2011}), \bibinfo{pages}{51--58}.
\newblock


\bibitem[\protect\citeauthoryear{Ronen, Shamir, Weingarten, and
  O’Flynn}{Ronen et~al\mbox{.}}{2017}]%
        {ronen2017iot}
\bibfield{author}{\bibinfo{person}{Eyal Ronen}, \bibinfo{person}{Adi Shamir},
  \bibinfo{person}{Achi-Or Weingarten}, {and} \bibinfo{person}{Colin
  O’Flynn}.} \bibinfo{year}{2017}\natexlab{}.
\newblock \showarticletitle{IoT goes nuclear: Creating a ZigBee chain
  reaction}. In \bibinfo{booktitle}{\emph{Security and Privacy (SP), 2017 IEEE
  Symposium on}}. IEEE, \bibinfo{pages}{195--212}.
\newblock


\bibitem[\protect\citeauthoryear{Ryan}{Ryan}{2013}]%
        {ryan2013bluetooth}
\bibfield{author}{\bibinfo{person}{Mike Ryan}.}
  \bibinfo{year}{2013}\natexlab{}.
\newblock \showarticletitle{Bluetooth: With low energy comes low security}. In
  \bibinfo{booktitle}{\emph{Presented as part of the 7th $\{$USENIX$\}$
  Workshop on Offensive Technologies}}.
\newblock


\bibitem[\protect\citeauthoryear{Sandhya and Devi}{Sandhya and Devi}{2012}]%
        {sandhya2012analysis}
\bibfield{author}{\bibinfo{person}{S Sandhya} {and}
  \bibinfo{person}{KA~Sumithra Devi}.} \bibinfo{year}{2012}\natexlab{}.
\newblock \showarticletitle{Analysis of Bluetooth threats and v4. 0 security
  features}. In \bibinfo{booktitle}{\emph{2012 International Conference on
  Computing, Communication and Applications}}. IEEE, \bibinfo{pages}{1--4}.
\newblock


\bibitem[\protect\citeauthoryear{Santoso and Vun}{Santoso and Vun}{2015}]%
        {santoso2015securing}
\bibfield{author}{\bibinfo{person}{Freddy~K Santoso} {and}
  \bibinfo{person}{Nicholas~CH Vun}.} \bibinfo{year}{2015}\natexlab{}.
\newblock \showarticletitle{Securing IoT for smart home system}. In
  \bibinfo{booktitle}{\emph{Consumer Electronics (ISCE), 2015 IEEE
  International Symposium on}}. IEEE, \bibinfo{pages}{1--2}.
\newblock


\bibitem[\protect\citeauthoryear{Sawilla and Ou}{Sawilla and Ou}{2008}]%
        {Sawilla2008IdentifyingGraphs}
\bibfield{author}{\bibinfo{person}{Reginald~E Sawilla} {and}
  \bibinfo{person}{Xinming Ou}.} \bibinfo{year}{2008}\natexlab{}.
\newblock \showarticletitle{Identifying critical attack assets in dependency
  attack graphs}. In \bibinfo{booktitle}{\emph{European Symposium on Research
  in Computer Security}}. Springer, \bibinfo{pages}{18--34}.
\newblock


\bibitem[\protect\citeauthoryear{Singhal and Ou}{Singhal and Ou}{2011}]%
        {Singhal2011SecurityGraphs}
\bibfield{author}{\bibinfo{person}{Anoop Singhal} {and}
  \bibinfo{person}{Ximming Ou}.} \bibinfo{year}{2011}\natexlab{}.
\newblock \showarticletitle{{Security Risk Analysis of Enterprise Networks
  Using Probabilistic Attack Graphs}}.
\newblock \bibinfo{journal}{\emph{NIST Interagency Report 7788, National
  iInstitute of Standards and Technology, U.S. Department of Commerce.}}
  (\bibinfo{year}{2011}).
\newblock


\bibitem[\protect\citeauthoryear{Skarmeta, Hernandez-Ramos, and
  Moreno}{Skarmeta et~al\mbox{.}}{2014}]%
        {skarmeta2014decentralized}
\bibfield{author}{\bibinfo{person}{Antonio~F Skarmeta}, \bibinfo{person}{Jose~L
  Hernandez-Ramos}, {and} \bibinfo{person}{M~Victoria Moreno}.}
  \bibinfo{year}{2014}\natexlab{}.
\newblock \showarticletitle{A decentralized approach for security and privacy
  challenges in the internet of things}. In \bibinfo{booktitle}{\emph{Internet
  of Things (WF-IoT), 2014 IEEE World Forum on}}. IEEE,
  \bibinfo{pages}{67--72}.
\newblock


\bibitem[\protect\citeauthoryear{Swiler, Phillips, Ellis, and Chakerian}{Swiler
  et~al\mbox{.}}{2001}]%
        {swiler2001computer}
\bibfield{author}{\bibinfo{person}{Laura~P Swiler}, \bibinfo{person}{Cynthia
  Phillips}, \bibinfo{person}{David Ellis}, {and} \bibinfo{person}{Stefan
  Chakerian}.} \bibinfo{year}{2001}\natexlab{}.
\newblock \showarticletitle{Computer-attack graph generation tool}. In
  \bibinfo{booktitle}{\emph{Proceedings DARPA Information Survivability
  Conference and Exposition II. DISCEX'01}}, Vol.~\bibinfo{volume}{2}. IEEE,
  \bibinfo{pages}{307--321}.
\newblock


\bibitem[\protect\citeauthoryear{Vaccari, Cambiaso, and Aiello}{Vaccari
  et~al\mbox{.}}{2017}]%
        {Vaccari2017RemotelyNetworks}
\bibfield{author}{\bibinfo{person}{Ivan Vaccari}, \bibinfo{person}{Enrico
  Cambiaso}, {and} \bibinfo{person}{Maurizio Aiello}.}
  \bibinfo{year}{2017}\natexlab{}.
\newblock \showarticletitle{{Remotely Exploiting at Command Attacks on ZigBee
  Networks}}.
\newblock \bibinfo{journal}{\emph{Security and Communication Networks}}
  \bibinfo{volume}{2017} (\bibinfo{year}{2017}).
\newblock
\showISSN{19390122}
\urldef\tempurl%
\url{https://doi.org/10.1155/2017/1723658}
\showDOI{\tempurl}


\bibitem[\protect\citeauthoryear{Wang, Islam, Long, Singhal, and Jajodia}{Wang
  et~al\mbox{.}}{2008}]%
        {Wang2008AnMetric}
\bibfield{author}{\bibinfo{person}{Lingyu Wang}, \bibinfo{person}{Tania Islam},
  \bibinfo{person}{Tao Long}, \bibinfo{person}{Anoop Singhal}, {and}
  \bibinfo{person}{Sushil Jajodia}.} \bibinfo{year}{2008}\natexlab{}.
\newblock \showarticletitle{{An attack graph-based probabilistic security
  metric}}.
\newblock \bibinfo{journal}{\emph{Lecture Notes in Computer Science (including
  subseries Lecture Notes in Artificial Intelligence and Lecture Notes in
  Bioinformatics)}}  \bibinfo{volume}{5094 LNCS} (\bibinfo{year}{2008}),
  \bibinfo{pages}{283--296}.
\newblock
\showISBNx{354070566X}
\showISSN{03029743}
\urldef\tempurl%
\url{https://doi.org/10.1007/978-3-540-70567-3{\_}22}
\showDOI{\tempurl}


\bibitem[\protect\citeauthoryear{Wright}{Wright}{2009}]%
        {wright2009killerbee}
\bibfield{author}{\bibinfo{person}{Joshua Wright}.}
  \bibinfo{year}{2009}\natexlab{}.
\newblock \showarticletitle{Killerbee: practical zigbee exploitation
  framework}. In \bibinfo{booktitle}{\emph{11th ToorCon conference, San
  Diego}}, Vol.~\bibinfo{volume}{67}.
\newblock


\bibitem[\protect\citeauthoryear{Yi{\u{g}}it, G{\"u}r, Alag{\"o}z, and
  Tellenbach}{Yi{\u{g}}it et~al\mbox{.}}{2019}]%
        {yiugit2019cost}
\bibfield{author}{\bibinfo{person}{Beyt{\"u}llah Yi{\u{g}}it},
  \bibinfo{person}{G{\"u}rkan G{\"u}r}, \bibinfo{person}{Fatih Alag{\"o}z},
  {and} \bibinfo{person}{Bernhard Tellenbach}.}
  \bibinfo{year}{2019}\natexlab{}.
\newblock \showarticletitle{Cost-aware securing of IoT systems using attack
  graphs}.
\newblock \bibinfo{journal}{\emph{Ad Hoc Networks}}  \bibinfo{volume}{86}
  (\bibinfo{year}{2019}), \bibinfo{pages}{23--35}.
\newblock


\bibitem[\protect\citeauthoryear{Yu, Sekar, Seshan, Agarwal, and Xu}{Yu
  et~al\mbox{.}}{2015}]%
        {Yu2015HandlingDevices}
\bibfield{author}{\bibinfo{person}{Tianlong Yu}, \bibinfo{person}{Vyas Sekar},
  \bibinfo{person}{Srinivasan Seshan}, \bibinfo{person}{Yuvraj Agarwal}, {and}
  \bibinfo{person}{Chenren Xu}.} \bibinfo{year}{2015}\natexlab{}.
\newblock \showarticletitle{{Handling a trillion (unfixable) flaws on a billion
  devices}}. In \bibinfo{booktitle}{\emph{Proceedings of the 14th ACM Workshop
  on Hot Topics in Networks - HotNets-XIV}}.
\newblock
\showISBNx{9781450340472}
\urldef\tempurl%
\url{https://doi.org/10.1145/2834050.2834095}
\showDOI{\tempurl}


\bibitem[\protect\citeauthoryear{Zanella, Bui, Castellani, Vangelista, and
  Zorzi}{Zanella et~al\mbox{.}}{2014}]%
        {zanella2014internet}
\bibfield{author}{\bibinfo{person}{Andrea Zanella}, \bibinfo{person}{Nicola
  Bui}, \bibinfo{person}{Angelo Castellani}, \bibinfo{person}{Lorenzo
  Vangelista}, {and} \bibinfo{person}{Michele Zorzi}.}
  \bibinfo{year}{2014}\natexlab{}.
\newblock \showarticletitle{Internet of things for smart cities}.
\newblock \bibinfo{journal}{\emph{IEEE Internet of Things journal}}
  \bibinfo{volume}{1}, \bibinfo{number}{1} (\bibinfo{year}{2014}),
  \bibinfo{pages}{22--32}.
\newblock


\bibitem[\protect\citeauthoryear{Zhang, Cho, Wang, Hsu, Chen, and Shieh}{Zhang
  et~al\mbox{.}}{2014}]%
        {zhang2014iot}
\bibfield{author}{\bibinfo{person}{Zhi-Kai Zhang}, \bibinfo{person}{Michael
  Cheng~Yi Cho}, \bibinfo{person}{Chia-Wei Wang}, \bibinfo{person}{Chia-Wei
  Hsu}, \bibinfo{person}{Chong-Kuan Chen}, {and} \bibinfo{person}{Shiuhpyng
  Shieh}.} \bibinfo{year}{2014}\natexlab{}.
\newblock \showarticletitle{IoT security: ongoing challenges and research
  opportunities}. In \bibinfo{booktitle}{\emph{Service-Oriented Computing and
  Applications (SOCA), 2014 IEEE 7th International Conference on}}. IEEE,
  \bibinfo{pages}{230--234}.
\newblock


\bibitem[\protect\citeauthoryear{Zhou and Hansen}{Zhou and Hansen}{2006}]%
        {Zhou2006Breadth}
\bibfield{author}{\bibinfo{person}{Rong Zhou} {and} \bibinfo{person}{Eric~A.
  Hansen}.} \bibinfo{year}{2006}\natexlab{}.
\newblock \showarticletitle{{Breadth-first heuristic search}}.
\newblock \bibinfo{journal}{\emph{Artificial Intelligence}}
  (\bibinfo{year}{2006}).
\newblock
\showISBNx{00043702}
\showISSN{00043702}
\urldef\tempurl%
\url{https://doi.org/10.1016/j.artint.2005.12.002}
\showDOI{\tempurl}


\bibitem[\protect\citeauthoryear{Zillner}{Zillner}{2015}]%
        {Zillner2015ZigBeeUgly}
\bibfield{author}{\bibinfo{person}{Tobias Zillner}.}
  \bibinfo{year}{2015}\natexlab{}.
\newblock \showarticletitle{{ZigBee Exploited - The Good, the Bad and the
  Ugly}}.
\newblock \bibinfo{journal}{\emph{Black Hat}} \bibinfo{volume}{16},
  \bibinfo{number}{2} (\bibinfo{year}{2015}), \bibinfo{pages}{6}.
\newblock
\urldef\tempurl%
\url{https://doi.org/10.1007/s11229-007-9263-9}
\showDOI{\tempurl}


\end{thebibliography}

\end{document}